\documentstyle[12pt]{article}
\begin{document}


\sloppy
\title
 {
      \hfill{\normalsize\sf FIAN/TD/00-01}    \\
            \vspace{1cm}
{\large Once again on the equivalence theorem. }
 }
\author
 {
       I.V.Tyutin
          \thanks
             {E-mail: tyutin@td.lpi.ac.ru}
  \\
               {\small \phantom{uuu}}
  \\
           {\it {\small} I.E.Tamm Department of Theoretical Physics,}
  \\
               {\it {\small} P.N.Lebedev Physical Institute,}
  \\
         {\it {\small} 117924, Leninsky Prospect 53, Moscow, Russia.}
 }
\date{ }
\maketitle
\begin{abstract}
We present the proof of the equivalence theorem in quantum field
theory which is based on a formulation of this problem in the field-anitfield
formalism.  As an example, we consider a model in which a different choices
of natural finite counterterms is possible, leading to physically non-equivalent
quantum theories while the equivalent theorem remains valid.
\end{abstract}

\newpage

\section{Introduction}

The equivalence theorem in the Lagrangian quantum field theory (that states the
independence of physical observables, in particular, the $S$--matrix, in
quantum theory on changes of variables in the classical Lagrangian, i.e. on
the choice of parametrization of the classical action) has a long story
\cite{Dys} -- \cite{BogLogOksTod}. The first rigorous result is due to
Borchers \cite{Bor} who proves that the $S$--matrix for the field that is the
local normally ordered polinomial of a free field and has non--zero
$in$--limit coincides with the $S$--matrix for the free field, i.e. is equal
to unity (on the generalization of Borchers's results for theories with
interaction in the framework of the axiomatic approach see
\cite{BogLogOksTod}). For the theories with non--zero interaction the
rigorous perturbative proof of the equivalence was given in \cite{Lam},
\cite{BerLam}. In these papers, the quantum action principle was used in the
form that coincides with the formal expression following, for example, from
formal manipulations with the functional integral representation for the
Green functions. However, in the general case, the form of quantum action
principle differs from the formal expression by the so called local
insertions (see sect. 3). In this paper, we present the perturbative proof of
the equivalence theorem that is valid for any quantum theory renormalized
with the use of the Bogoliubov $R$--operation \cite{BogShir} (see also
\cite{Zav} and reference therein). The changes of variables in classical
action are treated as specific symmetries of this action. The problem of the
proof of the equivalence theorem reduces to the problem of the possibility to
conserve this symmetry in the quantum theory. To solve this problem, we use
the generalization of the field--antifield formalism \cite{BV} to the case of
global symmetries \cite{VT7} -- \cite{BraHenWil2} and the cogomological
method for studying the symmetry structure in the renormalized theory, this
method was successfully applied to gauge theories (see \cite{PigSor},
\cite{BarBraHen} and reference therein). We show that the equivalence theorem
is valid in the sense that we can always choose the finite quantum corrections
(dependent on the action parametrization) to the classical action such that
the physical observables and $S$--matrix do not depend on the choice of the
classical action parametrization. There is a rather large arbitrariness in
the choice of the finite counterterms such that we can obtain the physically
nonequivalent families of quantum theories, the equivalence theorem being
valid inside each of these. The paper is organized as follows. In Section 2,
we present the formal considerations of the equivalence theorem. In Section
3, we derive the basic equation for the vertex function generating
functional, from which the equivalence theorem follows. In Section 4 we
consider the example of the theory where the different natural choice of
counterterms is possible that leads to the physically nonequivalent theories
without breaking the equivalence theorem.

\section{Formal consideration}

In this section, we briefly recall the must convenient for our purposes
scheme of proving the equivalence theorem. Let
$$
S_0=S_0(\varphi)=\int dxL(\varphi^i(x),\partial_\mu\varphi^i(x),\ldots)
$$
be a classical action. For simplicity, we assume that all the fields (which
we also call variables) $\varphi^i(x)\equiv\varphi^A$ are the Bose ones,and
the Lagrangian density $L$ depends on a finite number of the derivatives of
the fields $\varphi^i(x)$ (at least, perturbatively). We consider the family
of classical actions $S(\alpha,\varphi)$:
\begin{equation}\label{Act}
S(\alpha,\varphi)=S_0(\Phi(\alpha,\varphi)),
\end{equation}
where the change of variables $\Phi^A(\alpha,\varphi)=\Phi^i(\alpha,\varphi;x)
=\Phi^i(\alpha,\varphi^j(x),\partial_\mu\varphi^j(x),\ldots)=\varphi^A+O(g)$
($g$ denotes the total set of the coupling constants of the theory),
$\Phi^i(0,\varphi;x)=\varphi^i(x)$, and its inverse are local (at least,
perturbatively); the quantities
$$
f^A=f^A(\alpha,\varphi)={\partial\Phi^B(\alpha,\varphi)\over\partial\alpha}
{\delta\varphi^A(\alpha,\Phi)\over\delta\Phi^B},\quad
\varphi^A(\alpha,\Phi(\alpha,\varphi))= \varphi^A,
$$
being (perturbatively) local functions of $\varphi^i(x)$. We obviously have
$$
S(0,\varphi)=S_0(\varphi).
$$

The fact that the action $S(\alpha,\varphi)$ is obtained from the action
$S_0(\varphi)$ by the change of variables leads to the (symmetry) equation
for $S(\alpha,\varphi)$:
\begin{equation}\label{Sym}
{\partial S(\alpha,\varphi)\over\partial\alpha}-
f^A(\alpha,\varphi){\delta S(\alpha,\varphi)\over\delta\varphi^A}=0.
\end{equation}
Eq. (\ref{Sym}) is satisfied because the change of variables obeys this
equation:
$$
{\partial\Phi^A(\alpha,\varphi)\over\partial\alpha}-
f^B(\alpha,\varphi){\delta\Phi^A(\alpha,\varphi)\over\delta\varphi^B}=0.
$$

Classical theories related by a change of variables are equivalent. Naively,
the same situation takes places in quantum field theory. Consider the family
of the Green function generating functionals $Z(\alpha,J)$:
$$
Z(\alpha,J)=e^{iW(\alpha,J)}=\int
D\varphi\Delta(\alpha,\varphi)
e^{i\left(S(\alpha,\varphi)+J_A\varphi^A\right)},
$$
$$
\Delta(\alpha,\varphi)=
\hbox{Det}{\delta\Phi^A(\alpha,\varphi)\over\delta\varphi^B}.
$$
Evaluating the mean of symmetry equation (\ref{Sym}) and formally integrating
by parts, we obtain the equation for the Green function generating
functionals:
\begin{equation}\label{Green}
{\partial\over\partial\alpha}W(\alpha,J)+J^A\langle f^A\rangle(\alpha,J)=0,
\end{equation}
$$
\langle f^A\rangle(\alpha,J)\equiv {1\over Z(\alpha,J)}\int D\varphi
f^A(\alpha,\varphi)\Delta(\alpha,\varphi)
e^{i\left(S(\alpha,\varphi)+J_A\varphi^A\right)},
$$
or, equivalently, the equation for the vertex function generating
functional:
\begin{equation}\label{Vert}
{\partial\Gamma(\alpha,\varphi)\over\partial\alpha}-
\langle f^A\rangle(\alpha,J(\alpha,\varphi))
{\delta\Gamma(\alpha,\varphi)\over\delta\varphi^A}=0,
\end{equation}
$$
\Gamma(\alpha,\varphi)=W(\alpha,J)-J_A\varphi^A,\quad \varphi^A=
{\delta W(\alpha,J)\over\delta J_A},
$$
$J$ in the functional $\langle f^A\rangle$ of equation (\ref{Vert}) being
expressed in terms $\varphi$. An equation of the type (\ref{Vert}) for the
vertex function generating functional will be called the basic equation.

We assume that one--particle irreducible components of skeleton diagrams
(i.e. one--particle irreducible skeleton subdiagrams that are not contained
in any other one--particular irreducible skeleton subdiagrams) have no
one--particle pole singularities with respect to momentum conjugated to the
coordinates of the vertex $f^i(\alpha,\varphi;x)$ (this assumption is
certainly valid if all the fields are massive).  Then it follows from eqs
(\ref{Green}) or (\ref{Vert}) (see, for example \cite{KalT}), that the
masses of particles and the $S$--matrix elements do not depend on $\alpha$.

The deficiency of this consideration is that none of used quantities of
quantum field theory ($Z$, $W$, $\Gamma$, $\langle f^A\rangle$) does not
exist because of the known ultraviolet divergencies. We can however make a
useful conclusion from this formal consideration. Really, the equivalences
theorem is based on the equation of the type (\ref{Green}) or (\ref{Vert}).
If we establish that finite (renormalized) generating functionals satisfy the
equations of the type (\ref{Green}) or (\ref{Vert}), where
$\langle f^A\rangle$ is the mean of local operator, this will imply... that
masses and $S$--matrix elements do not depend on $\alpha$, which we interpret
as the equivalence theorem. In the next section, we show that in any theory
that can be made finite by a renormalization of the Bogoliubov $R$--operation
type, we can succeed for the vertex function generating functional to satisfy
the basic equation of the type (\ref{Vert}).

\section{Basic equation}

As we said above, the fact that the action $S(\alpha,\varphi)$ is obtained
from the action $S_0(\varphi)$ by the change of variables can be treated as
the presence of the global symmetry of the action $S(\alpha,\varphi)$,
whose infinitesimal form is
$$
\delta\varphi^A=-f^A\theta, \quad \delta\alpha=\theta,
$$
$\theta$ is a parameter of the global symmetry transformation. To study global
symmetries in quantum field theory, it is convenient to use the
field--antifield formalism \cite{VT7} -- \cite{BraHenWil2} developed by
Batalin and Vilkovisky for local (gauge) symmetries \cite{BV}. We shall
follow this strategy.

In accordance with the presence of the global symmetry, we introduce an
additional global ghost variable $c$, $\varepsilon(c)=1$, $c^2=0$, and the
antivariables $\varphi^*_i(x)$ with the opposite Grassman parity associated
with variables $\varphi^i(x)$ (we do not need antivariables $\alpha^*$,
$c^*$). We assign a ghost number $gh$ to every variables:
$$
gh(\varphi^A)=gh(\alpha)=0,\quad gh(\varphi^*_A)=-1, \quad gh(с)=1.
$$
In what follows, the total set of variables is denoted by $\eta$:
$\eta=\{\varphi^A,\varphi^*_A, \alpha,c\}$, the set of variables $\varphi^A$,
$\alpha$ is denoted by $\xi$:  $\xi=\{\varphi^A,\alpha\}$, the dependence on
these variables being explicitly indicated.

We take the master action ${\cal S}(\eta)$, $\varepsilon({\cal S})=gh(S)=0$,
to be:
$$
{\cal S}(\eta)=S(\xi)-\varphi^*_Af^A(\xi)c.
$$
$S(\eta)$ satisfies the master equation
\begin{equation}\label{Mast}
{1\over2}({\cal S}(\eta),{\cal S}(\eta))+
c{\partial{\cal S}(\eta)\over\partial\alpha}
=\left({\partial S(\xi)\over\partial\alpha}-
f^A(\xi){\delta S(\xi)\over\delta\varphi^A}\right)c=0,
\end{equation}
where the antibracket $(F,G)$ for functionals $F$ and $G$ is defined as:
$$
(F,G)=F{\overleftarrow\delta\over\delta\varphi^A}
{\delta\over\delta\varphi^*_A}G-
F{\overleftarrow\delta\over\delta\varphi^*_A}{\delta\over\delta\varphi^A}G.
$$
The only consequence of master equation (\ref{Mast}) is eq. (\ref{Sym}) for
the functional $S(\eta)\biggr|_{\varphi^*=0}=S(\xi)$, whose general solution
is
$$
S(\xi)=S_0(\Phi(\xi))
$$
with some functional $S_0$. So, if $S(\xi)$ has form (\ref{Act}), the master
action ${\cal S}(\eta)$ satisfies the master equation. Inversely, if we
require for the master action ${\cal S}(\eta)$ to satisfy master equation
(\ref{Mast}), then the action $S(\xi)$ will have the form (\ref{Act}).

The Green function generating functional is defined by
$$
Z(J)=e^{{i\over\hbar}W(J)}=\langle1\rangle,
$$
$$
\langle Q\rangle\equiv\langle Q(\eta)e^{{i\over\hbar}({\cal S}_{int}(\eta)+
J_A\varphi^A)}\rangle_{ren}, $$ $$ {\cal S}_{int}(\eta)={\cal S}(\eta)-
S_2(\varphi),
$$
where $S_2(\varphi)={\cal S}(\eta)|_{g=\varphi^*=0}$, $Q(\eta)$ is an
arbitrary functional, and $\langle(\ldots)\rangle_{ren}$ implies the mean
over the free vacuum of the expression in the parenthesis using the Feynman
rules with the free propagators defined by action $S_2(\varphi)$ and some
regularization and renormalization procedure. We do not need an explicit form
of the regularization scheme, however we assume that the quantum action
principle is valid for the finite Green functions (see \cite{PigSor} and
reference therein; all the scheme used at present satisfy this assumption).
In particular, the following properties are valid (${\cal T}(\eta)$ is the
vertex function generating functional for the theory with action
${\cal S}(\eta)$):

(i)
$$
{\partial\over\partial\lambda}{\cal T}(\eta)=
Q(\hbar,\eta)\circ{\cal T}(\eta), \quad
Q(\hbar,\eta)={\partial\over\partial\lambda}{\cal S}(\eta)+ \hbar
Q^{(1)}(\hbar,\eta),
$$
where the operation (the so called insertion)
$Q(\hbar,\eta)\circ{\cal T}(\eta)$ implies that the vertex function generating
functional are evaluated in accordance with the standard Feynman rules with
the additional vertex $Q(\hbar,\eta)$, $\lambda$ is an arbitrary parameter of
the theory under consideration, and $Q^{(1)}(\hbar,\eta)$ is some local
functional, which is equal to zero if parameter $\lambda$ does not appear in
the free propagators (i.e. it enters only  ${\cal S}_{int}$).

(ii)
$$
{\delta\over\delta\varphi^A}{\cal T}(\eta)=
Q_A(\hbar,\eta)\circ{\cal T}(\eta),
\quad Q_A(\hbar,\eta)={\delta\over\delta\varphi^A}{\cal S}(\eta)+
\hbar Q^{(1)}_A(\hbar,\eta),
$$
where $Q^{(1)}_A(\hbar,\eta)$ are some local functionals.

(iii) The vertex function generating functional ${\cal T}(\eta)$ is Poincare
invariant and has all the linear homogeneous symmetries of the action
functional ${\cal S}(\eta)$ that do not touch the space--time coordinates and
Lorentzian indices. In particular, in the case under consideration, the
vertex function generating functional conserves the ghost number.

The regularization properties (i), (ii) enable us to establish \cite{PigSor}
that the vertex function generating functional satisfies eq. (\ref{Mast}) up
to the local insertions:
\begin{equation}\label{TW}
{1\over2}({\cal T}(\eta),{\cal T}(\eta))+
c{\partial{\cal T}(\eta)\over\partial\alpha}=
-\hbar Q^{(1)}(\hbar,\eta)\circ{\cal T}(\eta),
\end{equation}
$$
Q^{(1)}(\hbar,\eta)=Q^{(1)}_0(\eta)+O(\hbar).
$$
The local insertions must satisfy the equation that is the consistency
condition for eq. (\ref{TW}):
$$
({\cal T}(\eta),Q^{(1)}(\hbar,\eta)\circ{\cal T}(\eta))+
c{\partial\over\partial\alpha}(Q^{(1)}(\hbar,\eta)\circ{\cal T}(\eta))=0.
$$
We have in the one--loop approximation
$$
({\cal T}_{[1]}(\eta),{\cal T}_{[1]}(\eta))_{[1]}=-\hbar Q^{(1)}_0(\eta),
$$
$$
{\cal T}(\eta)={\cal S}(\eta)+\hbar{\cal T}_1(\eta)+O(\hbar^2),
$$
and the lower index ``$[n]$'' at any functional $G$ implies that only the
first $n+1$ terms of the Teylor series in $\hbar$ are taken into account:
$$
G\equiv G_{[n]}+O(\hbar^{n+1}),\quad
{\partial^{n+1}\over\partial\hbar^{n+1}}G_{[n]}=0.
$$

Because of the ghost number conservation, the functional $Q^{(1)}_0(\eta)$
has a ghost number 1, therefore, it is linear in $c$ and does not depend on
$\varphi^*_A$:
$$
Q^{(1)}_0(\eta)=cq^{(1)}(\xi).
$$
The consistency condition in the one--loop approximation
$$
\omega Q^{(1)}_0(\eta)=0,
$$
$$
\omega={\delta{\cal S}(\eta)\over\delta\varphi^A}
{\delta\over\delta\varphi^*_A}+
{\delta{\cal S}(\eta)\over\delta\varphi^*_A}{\delta\over\delta\varphi^A}
+c{\partial\over\partial\alpha},\quad \omega^2=0,
$$
is identically satisfied.

Lemma:

The functional $Q^{(1)}_0$ can be represented as
$$
Q^{(1)}_0(\eta)=\omega X^{(1)}(\eta)
$$
with some local functional $X^{(1)}(\eta)$, $gh(X^{(1)})=0$.

To prove the Lemma, it is convenient to pass from variables $\eta$ to
variables $\tilde{\eta}=\{\Phi^A,\Phi^*_A=
\varphi^*_B(\partial\varphi^B(\alpha,\Phi)/\partial\Phi^A),\alpha,c\}$.
With any functional $G(\eta)$, we also associate the functional
$\tilde{G}(\tilde{\eta})$:
$$
\tilde{G}(\tilde{\eta})=G(\eta(\tilde{\eta})),\quad
G(\eta)=\tilde{G}(\tilde{\eta}(\eta)).
$$
In this case
$$
\omega G(\eta)=\tilde{\omega}\tilde{G}(\tilde{\eta}),
$$
$$
\tilde{\omega}={\delta S_0(\Phi)\over\delta\Phi^A}
{\delta\over\delta\Phi^*_A}+c{\partial\over\partial\alpha}.
$$

In new variables, the statement of the Lemma considered as on $X^{(1)}(\eta)$
has the form:
\begin{equation}\label{Cogomol1}
\tilde{Q}^{(1)}_0(\tilde{\eta})=\tilde{\omega}\tilde{X}^{(1)}(\tilde{\eta}).
\end{equation}

To solve this equation , we introduce the operator
$\gamma=\alpha\partial/\partial c$. The operators $\omega$ and $\gamma$ forms
an algebra:
$$
\omega^2=\gamma^2=0,\quad
\omega\gamma+\gamma\omega=N,\quad [\omega,N]=[\gamma,N]=0,
$$
$$
N=\alpha{\partial\over\partial\alpha}+c{\partial\over\partial c}.
$$

A particular solution $\tilde{X}^{(1)}_p(\tilde{\eta})$ of (nonhomogeneous)
eq. (\ref{Cogomol1}) can be taken in the form:
$$
\tilde{X}^{(1)}_p(\tilde{\eta})=
{1\over N}\gamma\tilde{Q}^{(1)}_0(\tilde{\eta}),
$$
where the action of an arbitrary function $f(N)$ of operator $N$ is defined
by:
$$
f(N)\alpha^kc^l=f(k+l)\alpha^kc^l,\quad k,l\ge0.
$$
We note that $\tilde{X}^{(1)}_p(\tilde{\eta})$ does not depend on $\varphi^*$
and $c$: $\tilde{X}^{(1)}_p=\tilde{X}^{(1)}_p(\tilde{\xi})$. The general
solution to eq. (\ref{Cogomol1}) is obtained by adding the general solution
$\tilde{X}^{(1)}_h(\tilde{\eta})$ of the homogeneous equation
\begin{equation}\label{Cogomol2}
\tilde{\omega}\tilde{X}^{(1)}_h(\tilde{\eta})=0.
\end{equation}
to $\tilde{X}^{(1)}_p(\tilde{\xi})$. We present
$\tilde{X}^{(1)}_h(\tilde{\eta})$ as
$$
\tilde{X}^{(1)}_h(\tilde{\eta})=S_{01}(\Phi)+
\tilde{X}^{(1)}_{h1}(\tilde{\eta}),\quad S_{01}(\Phi)=
\tilde{X}^{(1)}_{h}(\tilde{\eta})|_{\alpha=c=0}
$$
($S_{01}$ depends only on $\Phi$ because $gh(\tilde{X}^{(1)}_h)=0$).
The functional $S_{01}(\Phi)$ do not enter eq. (\ref{Cogomol2}) and the
standard arguments give:
$$
\tilde{X}^{(1)}_{h1}(\tilde{\eta})=\tilde{\omega}
\tilde{Y}^{(1)}(\tilde{\eta}),\quad gh(\tilde{Y}^{(1)})=-1,
$$
with some local functional $\tilde{Y}^{(1)}(\tilde{\eta})=
\Phi^*_A\tilde{\Phi}^A_1(\tilde{\xi})$.

Returning to the initial variables, we obtain:
$$
X^{(1)}(\eta)=
$$
\begin{equation}\label{X1}
=S_{01}(\Phi)+{\delta S_0(\Phi)\over\delta\Phi^A}\Phi^A_1(\xi)-
\varphi^*_A{\partial\tilde{\varphi}^A\over\partial\Phi_B}
({\partial\Phi^B_1(\xi)\over\partial\alpha}-
{\partial\Phi^B_1(\xi)\over\partial\varphi^C}
{\partial\tilde{\varphi}^C\over\partial\Phi^D}
{\partial\Phi^D(\xi)\over\partial\alpha})c+X^{(1)}_p(\xi).
\end{equation}

We consider now the master action ${\cal S}^{(1)}(\eta)$:
$$
{\cal S}^{(1)}(\eta)={\cal S}(\eta)+\hbar X^{(1)}(\eta)\equiv S^{(1)}(\xi)-
\varphi^*_Af^{(1)A}(\xi)c,
$$
the function $f^{(1)A}(\xi)=f^A(\xi)+O(\hbar)$ being local. The sense of the
separate terms in expression (\ref{X1}) for $X^{(1)}$ becomes now clear: the
first term describes the quantum corrections to the initial classical
action, the second and third ones describe the quantum corrections to the
change of variables, the last term has to ``compensate'' a possible
noncovariance of the regularization scheme adopted.

The master action ${\cal S}^{(1)}(\eta)$ does not satisfy master equation
(\ref{Mast}):
$$
({\cal S}^{(1)}(\eta),{\cal S}^{(1)}(\eta))+
c{\partial{\cal S}^{(1)}(\eta)\over\partial\alpha}=
\hbar\Lambda^{(1)}(\hbar,\eta).
$$
It is however important, that $\Lambda^{(1)}(\hbar,\eta)$ is a local
functional.

The vertex function generating functional ${\cal T}^{(1)}(\eta)$ in the
theory with the action ${\cal S}^{(1)}(\eta)$ does satisfy the master
equation up to local insertion. But it is easy to verify that the local
insertions are absent in the one--loop approximation:
$$
{\cal T}^{(1)}_{[1]}(\eta)={\cal T}_{[1]}(\eta)+\hbar X^{(1)}(\eta),
$$
$$ {1\over2}({\cal T}^{(1)}_{[1]}(\eta),{\cal T}^{(1)}_{[1]}(\eta))_{[1]}
+c{\partial\over\partial\alpha}{\cal T}_{[1]}(\eta)= {1\over2}({\cal
T}_{[1]}(\eta),{\cal T}_{[1]}(\eta))_{[1]}
+c{\partial\over\partial\alpha}{\cal T}_{[1]}(\eta)+
$$
$$
+({\cal T}^{(1)}(\eta)_{[1]},\hbar X^{(1)}(\eta))_{[1]}+
\hbar c{\partial\over\partial\alpha}X^{(1)}(\eta)=
-\hbar Q^{(1)}_0(\eta)+\hbar\omega X^{(1)}(\eta)=0.
$$
So, the violation of the master equation for ${\cal T}^{(1)}(\eta)$
begins with the two--loop approximation:
$$
{1\over2}({\cal T}^{(1)}(\eta),{\cal T}^{(1)}(\eta))+
c{\partial\over\partial\alpha}{\cal T}^{(1)}(\eta)=
-\hbar^2Q^{(2)}(\hbar,\eta)\circ({\cal T}^{(1)}(\eta),
$$
$$
({\cal T}^{(1)}(\eta),Q^{(2)}(\hbar,\eta)\circ({\cal T}^{(1)}(\eta))+
c{\partial\over\partial\alpha}(Q^{(2)}(\hbar,\eta)\circ({\cal T}^{(1)}(\eta))=0.
$$

By induction, we finally obtain: There exists the action
${\cal S}^{(\infty)}(\eta)$,
$$
{\cal S}^{(\infty)}(\eta)={\cal S}(\eta)+
\sum\limits_{n=1}\hbar^nX^{(n)}(\eta)\equiv
S^{(\infty)}(\xi)-\varphi^*_Af^{(\infty)A}(\xi)c,
$$
where $f^{(\infty)A}(\xi)$ are local functions, such that the vertex function
generating functional ${\cal T}^{(\infty)}(\eta)$ satisfies the master
equation:
$$
{1\over2}({\cal T}^{(\infty)}(\eta),{\cal T}^{(\infty)}(\eta))+
c{\partial\over\partial\alpha}{\cal T}^{(\infty)}(\eta)=0.
$$
With the account of the relation (that holds in view of property (i)
of the regularizations used)
$$
{\delta\over\delta\varphi^*_A}{\cal T}^{(\infty)}(\eta)=
-c\langle f^{(\infty)A}\rangle^{(\infty)}(\xi),
$$
where the upper index ``$(\infty)$'' at the symbol of the mean implies that
the mean is calculated with the action $S^{(\infty)}(\xi)$, we obtain:
$$
{\partial\over\partial\alpha}\Gamma^{(\infty)}(\xi)-
\langle f^{(\infty)A}\rangle^{(\infty)}(\xi){\delta\over\delta\varphi^A}
\Gamma^{(\infty)}(\xi)=0,
$$
where $\Gamma^{(\infty)}(\xi)\equiv{\cal T}(\eta)|_{\varphi*=0}$ is the
vertex function generating functional for the theory with the
``renormalized'' action $S^{(\infty)}(\xi)$.

Thus, it is established that by adding the appropriate counterterms to the
initial action we can always make the vertex function generating functional
to satisfy basic equation (\ref{Vert}), i.e. the equivalence theorem to be
fulfilled.

\section{The example}

In this section, we are consider the example of the family of classical
theories related by the change of variables whose quantization leads to the
physically nonequivalent theories, the equivalence theorem being valid.

The model is described by action
\begin{equation}\label{Action}
S(\alpha,\psi)=S_0(\Psi(\alpha,\psi))=\bar{\psi}(i\gamma^\mu\partial_\mu+
\gamma^\mu V_\mu+\gamma^\mu\gamma^5A_\mu+
\alpha{f_\pi\over m}\gamma^\mu\gamma^5\partial_\mu\varphi)\psi,
\end{equation}
$$
S_0(\psi)=\bar{\psi}(i\gamma^\mu\partial_\mu+
\gamma^\mu V_\mu+\gamma^\mu\gamma^5A_\mu)\psi,\quad
\Psi(\alpha,\psi)=e^{-i\alpha{f_\pi\over m}\varphi}\psi,
$$
where $\psi(x)$ is a quantum Dirac field, $V_\mu(x)$, $A_\mu(x)$,
$\varphi(x)$ are respectively external vector, axial and pseudoscalar fields,
$\gamma^5=i\gamma^0\gamma^1\gamma^2\gamma^3$, the metrics is $diag(+,-,-,-)$,
the symbol $\int dx$ is omitted.

The vertex function generating functional must satisfy the equation:
$$
{\partial\over\partial\alpha}\Gamma^{(1)}-
\langle f_\psi^{(1)}\rangle^{(1)}{\delta\over\delta\psi}\Gamma^{(1)}-
\langle f_{\bar{\psi}}^{(1)}\rangle^{(1)}{\delta\over\delta\bar{\psi}}
\Gamma^{(1)}=0,
$$
where the upper index ``$(1)$'' means that the theory is exhausted by the
one-loop approximation,
$f_\psi^{(1)}=(-i{f_\pi\over m}\varphi\gamma^5+O(\hbar))\psi$,
$f_{\bar{\psi}}^{(1)}=\bar{\psi}(-i{f_\pi\over m}\varphi\gamma^5+O(\hbar))$,

We restrict ourselves to the discussion of vacuum diagrams, i.e. of the
vertex function generating functional for zero arguments $\psi$ и $\bar{\psi}$:
$$
\bar{\Gamma}\equiv\Gamma|_{\psi=\bar{\psi}=0}=
\bar{\Gamma}(\alpha,V_\mu,A_\mu,\varphi).
$$
In this limit $f_\psi^{(1)}=f_{\bar{\psi}}^{(1)}=0$, so that
$\bar{\Gamma}$ must satisfy the equation
\begin{equation}\label{Symm1}
{\partial\over\partial\alpha}\bar{\Gamma}=0.
\end{equation}

For $\varphi=0$ the expression $\tilde{\Gamma}(V_\mu,A_\mu)=
\bar{\Gamma}|_{\varphi=0}$ is uniquely defined by the requirement of the
exact conservation of the vector current:
$$
\partial_\mu{\delta\over\delta V_\mu(x)}\tilde{\Gamma}=0,
$$
and the conservation of the axial current, excepting the diagrams with
three external lines.  In this case
$$
\partial_\mu{\delta\over\delta A_\mu(x)}\tilde{\Gamma}=-{\hbar\over4\pi^2}
(\varepsilon^{\mu\nu\lambda\sigma}
\partial_\mu V_\nu(x)\partial_\lambda V_\sigma(x)
+{1\over3}\varepsilon^{\mu\nu\lambda\sigma}
\partial_\mu A_\nu(x)\partial_\lambda A_\sigma(x)),
$$
$\varepsilon^{0123}=1$. For $\varphi\neq0$ the expression for $\bar{\Gamma}$
is derived from the expression for  $\tilde{\Gamma}$ by the substitution
of $A_\mu+\alpha{f_\pi\over m}\partial_\mu\varphi$ for  $A_\mu$ and by the
addition of possible local counterterms (the conservation of the vector
current is required still):
$$
\bar{\Gamma}(\alpha,V_\mu,A_\mu,\varphi)=
\tilde{\Gamma}(V_\mu,A_\mu+\alpha{f_\pi\over m}\partial_\mu\varphi)+
\hbar S_{contr}(\alpha,V_\mu,A_\mu,\varphi).
$$
The dependence of $\tilde{\Gamma}$ on $\varphi$ may be calculated explicitly (for
example, through a differentiation by $\alpha$):
$$
\tilde{\Gamma}(V_\mu,A_\mu+\alpha{f_\pi\over m}\partial_\mu\varphi)=
$$
$$
=\tilde{\Gamma}(V_\mu,A_\mu)+{\alpha\hbar\over4\pi^2}{f_\pi\over m}
\int dx\left(\varphi(x)\varepsilon^{\mu\nu\lambda\sigma}
\partial_\mu V_\nu(x)\partial_\lambda V_\sigma(x)
+{1\over3}\varphi(x)\varepsilon^{\mu\nu\lambda\sigma}
\partial_\mu A_\nu(x)\partial_\lambda A_\sigma(x)\right).
$$

As for $S_{contr}(\alpha,V_\mu,A_\mu,\varphi)$, we shall only extract the term
linear in $\varphi$ and containing the tensor $\varepsilon^{\mu\nu\lambda\sigma}$.
The terms having other independent structure linear in $\varphi$ as well as the
terms of the qudaratic and higher powers in $\varphi$ are inessential for us:
$$
S_{contr}(\alpha,V_\mu,A_\mu,\varphi)=
$$
$$
={f_\pi\over m}\varphi\left(
r^\prime_1(\alpha)\varepsilon^{\mu\nu\lambda\sigma}
\partial_\mu V_\nu\partial_\lambda V_\sigma+
r^\prime_2(\alpha)\varepsilon^{\mu\nu\lambda\sigma}
\partial_\mu A_\nu\partial_\lambda A_\sigma\right)+
S^\prime_{contr}(\alpha,V_\mu,A_\mu,\varphi),
$$
where the sign of $\int dx$ is omitted.
Thus the general expression of $\bar{\Gamma}$ reads:
$$
\bar{\Gamma}(\alpha,V_\mu,A_\mu,\varphi)=\tilde{\Gamma}(V_\mu,A_\mu)+
$$
$$
+\hbar{f_\pi\over m}\varphi\left(
r_1(\alpha)\varepsilon^{\mu\nu\lambda\sigma}
\partial_\mu V_\nu\partial_\lambda V_\sigma+
r_2(\alpha)\varepsilon^{\mu\nu\lambda\sigma}
\partial_\mu A_\nu\partial_\lambda A_\sigma\right)+
\hbar S^\prime_{contr}(\alpha,V_\mu,A_\mu,\varphi),
$$
$$
r_1(\alpha)={\alpha\over4\pi^2}+r^\prime_1(\alpha),\quad
r_2(\alpha)={\alpha\over12\pi^2}+r^\prime_2(\alpha).
$$

The equation (\ref{Symm1}) is satisfied for the following choice of counterterms:
$$
r_1(\alpha)=-{\alpha\over4\pi^2}+r_1, \quad
r_2(\alpha)=-{\alpha\over12\pi^2}+r_2,\quad r_1,r_2=const,
$$
$$
S^\prime_{contr}=S^\prime_{contr}(V_\mu,A_\mu,\varphi),
$$
($r_1$, $r_2$ и $S^\prime_{contr}$ не зависят от $\alpha$).

As a result, we get the following expression for $\bar{\Gamma}$:
$$
\bar{\Gamma}(\alpha,V_\mu,A_\mu,\varphi)=\tilde{\Gamma}(V_\mu,A_\mu)+
$$
$$
+\hbar{f_\pi\over m}\varphi\left(r_1\varepsilon^{\mu\nu\lambda\sigma}
\partial_\mu V_\nu\partial_\lambda V_\sigma+
r_2\varepsilon^{\mu\nu\lambda\sigma}
\partial_\mu A_\nu\partial_\lambda A_\sigma\right)+
\hbar S^\prime_{contr}(V_\mu,A_\mu,\varphi).
$$

This expression clearly satisfies the equivalence theorem (it does not depend
on the change of variables in the classical action), however an ambiguity in
the choice of counterterms still remains.  This ambiguity could be explored
in different ways.

If one starts from a quantum theory which is constructed from the classical action
(\ref{Action}) при $\alpha=0$:
$$
S(0,\psi)=S_0(\psi))=\bar{\psi}(i\gamma^\mu\partial_\mu+
\gamma^\mu V_\mu+\gamma^\mu\gamma^5A_\mu)\psi,
$$
then it seems natural to require that for $\alpha=0$, and, consequently for
any $\alpha$ on the fermion mass shell the quantum theory does not depend on
the field $\varphi$.  This means that the functional $\bar{\Gamma}$ does not
depend on $\varphi$:
$$
\bar{\Gamma}=\tilde{\Gamma}(V_\mu,A_\mu).
$$

On the other hand, if one starts from the quantum theory which is
constructed from the classical action (\ref{Action}) for $\alpha=1$
(the choice of any other $\alpha\neq0$ as a normalization point
reduces to a redefinition of the parameter $f_\pi$ or $m$):
$$
S(1,\psi)=\bar{\psi}(i\gamma^\mu\partial_\mu+\gamma^\mu V_\mu+
\gamma^\mu\gamma^5A_\mu+
{f_\pi\over m}\gamma^\mu\gamma^5\partial_\mu\varphi)\psi,
$$
then it seems natural to demand that the field
${f_\pi\over m}\partial_\mu\varphi$
and the axial field $A_\nu$ should interact with the same
axial current.  In this case we must choose
$r_1=1/4\pi^2$, $r_2=1/12\pi^2$ (in addition, we put $S^\prime_{contr}=0$ for
simplicity):
$$
\bar{\Gamma}=\tilde{\Gamma}(V_\mu,A_\mu)+{\hbar\over4\pi^2}{f_\pi\over m}
\varphi\left(\varepsilon^{\mu\nu\lambda\sigma}
\partial_\mu V_\nu\partial_\lambda V_\sigma+
{1\over3}\varepsilon^{\mu\nu\lambda\sigma}
\partial_\mu A_\nu\partial_\lambda A_\sigma\right).
$$

Thus the example considered demonstrates that the requirement of validity
of the equivalence theorem does not eliminate the ambiguity related to
possible addition of finite counterterms.

{\bf Acknowledgments.} The work is  supported by Russian Foundation
for Basic Researches and by Human Capital and Mobility Program of the European
Community, grants RFBR--99--01--00980, RFBR--99--02--17916, INTAS--96--0308.

\end{document}